\documentclass[aps,twocolumn,floatfix,footinbib]{revtex4}
\usepackage[pdftex]{graphicx}
\usepackage{calc}
\usepackage[T1]{fontenc}
\usepackage[centertags,sumlimits]{amsmath}    
\usepackage{amsfonts}   
\usepackage{amssymb}    
\usepackage{bbm}
\usepackage[hang]{footmisc} 
\usepackage{textcomp} 
\usepackage{color}
\usepackage{setspace}
\usepackage[frenchb]{babel}
\begin{document}


\title{Hartree simulations of coupled quantum Hall edge states in corner-overgrown heterostructures}

\author{L. Steinke$^{1}$, P. Cantwell$^2$, 
E. Stach$^2$, D. Schuh$^{1,3}$, A. Fontcuberta i Morral$^{1,4}$, 
M. Bichler$^1$, G. Abstreiter$^1$, and M. Grayson$^{1,5\dagger}$}
\affiliation{$^1$Walter Schottky Institut, Technische Universit\"at M\"unchen, D-85748 
Garching, Germany}
\affiliation{$^2$School of Materials Engineering, Purdue University, West Lafayette, IN 47907, USA}
\affiliation{$^3$Laboratoire des Mat\'eriaux Semiconducteurs, Institut des Mat\'eriaux, EPFL, CH-1015 Lausanne, Switzerland}
\affiliation{$^4$Universit\"at Regensburg, Institut f\"ur Angewandte und Experimentelle Physik II,  D-93040 Regensburg, Germany}
\affiliation{$^5$Department of Electrical Engineering and Computer Science, Northwestern University, Evanston, IL 60208, USA}

\date{2 July 2010}

\begin{abstract}
The electronic states in a corner-overgrown bent GaAs/AlGaAs quantum well heterostructure are studied with numerical Hartree simulations. Transmission electron microscope pictures of the junction justify the sharp-corner assumption.  In a tilted magnetic field both facets of the bent quantum well are brought to a quantum Hall (QH) state, and the corner hosts an unconventional hybrid system of two coupled counter-propagating quantum Hall edges and an additional one-dimensional accumulation wire. A subsystems model is introduced, whereby the total hybrid dispersion and wavefunctions are explained in terms of the constituent QH edge- and accumulation wire-subsystem dispersions and wavefunctions.  At low magnetic fields, orthonormal basis wavefunctions of the hybrid system can be accurately estimated by projecting out the lowest bound state of the accumulation wire from the edge state wavefunctions.  At high magnetic fields, the coupling between the three subsystems {\em increases} as a function of the applied magnetic field, in contrast to coplanar barrier-junctions of QH systems, leading to large anticrossing gaps between the subsystem dispersions.  These results are discussed in terms of previously reported experimental data on bent quantum Hall systems.
\end{abstract} 
\maketitle
\section{Introduction}
The corner overgrowth technique \cite{2} realizes GaAs/AlGaAs heterostructures with two-dimensional (2D) layers bent at a sharp $90^{\circ}$ angle.
Magnetotransport measurements on a corner-overgrown bent GaAs/AlGaAs heterointerface structure \cite{4,thesis,3} have already demonstrated the high quality of these samples, showing electron mobilities of up to $2\times 10^{6}\,{\rm cm^2/Vs}$ at densities of order $10^{11}/{\rm cm}^2$. In a magnetic field $B$ these systems realize a unique sort of quantum Hall (QH) effect boundary state with either co- or counter-propagating one-dimensional (1D) edge modes coupled along the entire corner junction, at filling factors which depend on $B$-field tilt angle \cite{3}. At equal filling factors $\nu$ on both facets 
with counterpropagating edge modes, the system resembles a multimode 1D wire, whereby %
the 1D conductance along the corner junction exhibits 
strongly %
insulating, weakly insulating or metallic behavior, depending on $\nu$.   The metallic behavior represents a realization of the Kane-Fisher disordered anti-wire geometry for the fractional quantum Hall effect \cite{4}, 
whose hamiltonian is identical to that of a disordered 1D superconductor \cite{Kane,Renn}.%

Here we present numerical Hartree simulations of the 
dispersion, wavefunctions, and local %
carrier concentrations for bent quantum wells relevant to the transport samples of Refs. \cite{4,thesis,3}. 
The measurements show a corner profile of nanometer-scale sharpness similar to the diagnostic samples studied in Ref. \cite{TEM}. 
This paper 
first reviews %
the bent quantum well system at both zero magnetic field and at high fields of equal filling factor. In Section II, the structure will be reviewed and new 
transmission electron microscopy (TEM) %
pictures demonstrating the sharpness of the corner potential 
of an actual transport sample 
will be shown.  The 2D Hartree equation is solved in zero magnetic field in Section III, providing the Hartree potential used in subsequent sections. In Section IV, this 
Hartree solution 
will be solved 
in the presence of a weak magnetic field. 
To develop intuition for the resulting dispersion and wavefunctions in the low-$B$ limit, a simplified model is introduced for comparison
in Section V which separates the corner hamiltonian into an in-plane 1D sharp QH edge potential and an orthogonal 1D Hartree triangular confinement potential.   With this subsystems model,
%
the dispersions and wavefunctions of the constituent quantum Hall edges and the corner accumulation wire can be estimated.
It is shown that the %
2D %
Hartree solutions of the corner quantum well match closely to wavefunctions comprised of the constituent systems, provided that the sharp quantum Hall edge states have the deeply bound quantum wire state projected out. In the high $B$ limit, Section VI shows the resulting strong hybridization of the states, leading to dispersions with large gaps between hybridized Landau bands and regions of positive 
electron-like %
and negative 
hole-like %
curvature 
in the dispersion %
as each band traverses the corner. Concluding remarks in Section VII address how published experimental results can be interpreted with the microscopic understanding presented here.

\begin{figure}
\includegraphics[width=\columnwidth]{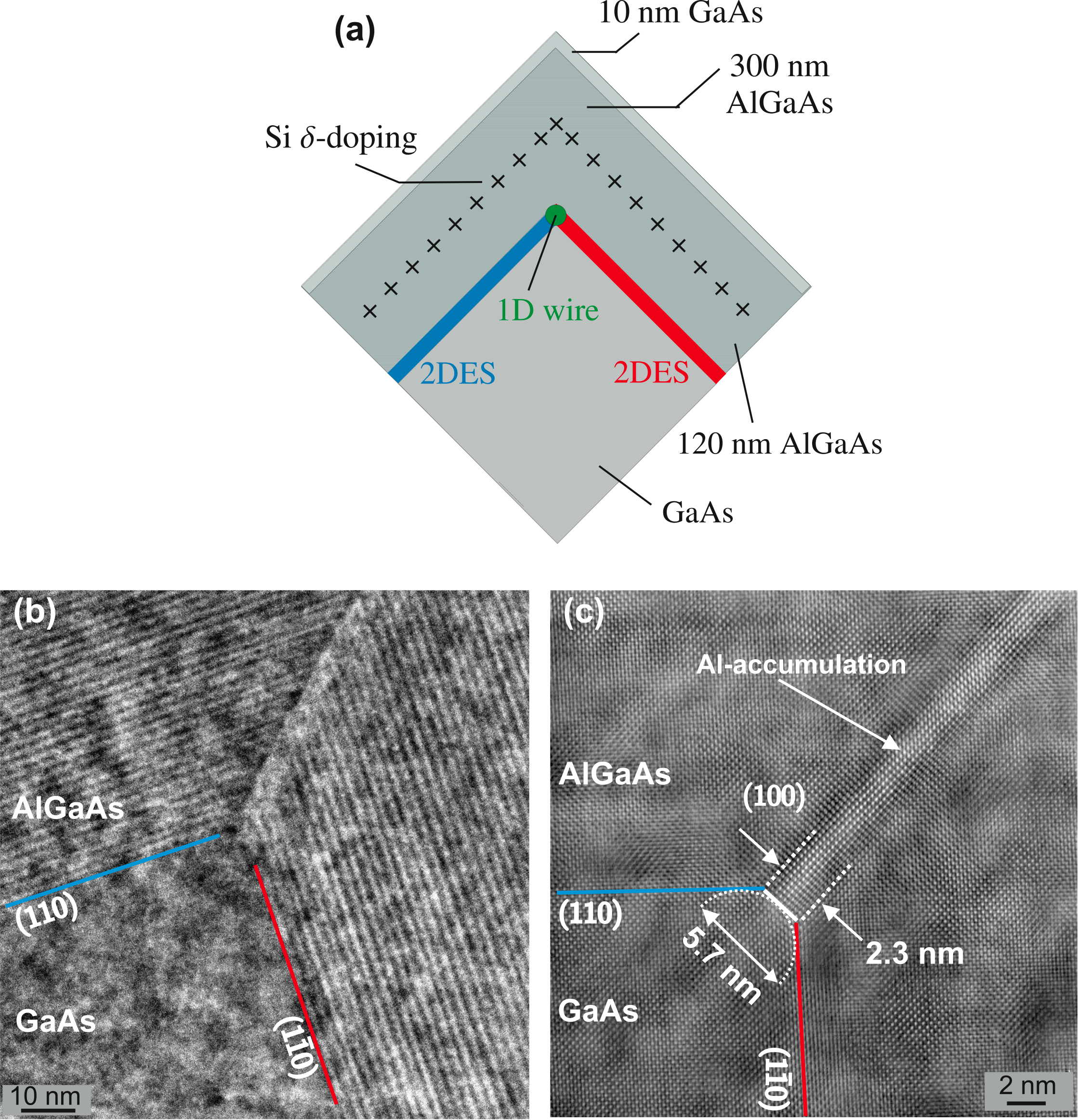}
\caption{(Color online) Panel (a) shows a schematic of the bent quantum well heterostructure. Electrons from the Si donors in the \mbox{$\delta$-doping} layer accumulate at the interface between the MBE-grown GaAs base layer and the AlGaAs spacer and form 2DES on the two facets (blue and red). At the corner an additional 1D charge accumulation arises (green). High-resolution bright-field TEM images in panels (b) and (c) show a cross-section of the bent heterointerface between the GaAs and AlGaAs
layer. Brighter regions correspond to a periodically higher Al content per period growth under rotation  \cite{TEM}.}%
\label{Fig1}
\end{figure}

\section{TEM and corner sharpness in the transport structure}
The growth technique and complete layer structure are presented in Refs.\,\cite{2,TEM}, with layer thicknesses summarized in Fig.\,\ref{Fig1} (a). The focus of this work is the electronic structure near the corner junction where electrons accumulate. A bent two-dimensional electron system (2DES) forms at the interface between an MBE grown GaAs base layer and the AlGaAs spacer.

The TEM images in panels (b) and (c) show a cross-section of the GaAs/AlGaAs interface at the overgrown corner, demonstrating a sharp corner profile with an effective diameter of curvature of \mbox
{$2r = 
5.7$ nm}. The high-contrast stripe along the diagonal in Fig.\,\ref{Fig1} (c) is an accumulation of Al adatoms, which occurs due to the slower diffusion of Al compared to Ga \cite{TEM}. 
Previously reported %
high-resolution TEM measurements were restricted to a 
test structure with 
high image-contrast 
GaAs/AlAs interfaces \cite{TEM}.
The images presented here are of an actual transport structure, and
discern for the first time that the transport layer of the %
GaAs/AlGaAs interface in corner-overgrown transport samples yield corner profiles with similar nanometer-scale sharpness.

The validity of the sharp-corner assumption for Hartree calculations depends on the corner curvature in relation to relevant quantum length scales \cite{TEM}. For example, the one-dimensional accumulation wire in Fig.\,\ref{Fig2} is predicted to exist at $B = 0$ in the corner-overgrown heterostructures if the diameter of curvature at the corner $2r$ is smaller than half the Fermi wavelength $\lambda_F $  \cite{TEM}. For typical sheet electron densities between $1.0\times10^{11}\,{\rm cm}^{-2}$ and $1.5\times10^{11}\,{\rm cm}^{-2}$ $\lambda_F$ is between 80 nm and 65 nm. With \mbox{$2r=5.7$ nm}, as evident from Fig.\,\ref{Fig1} (c), the condition $2r<\lambda_F/2$ is therefore satisfied, and a one-dimensional wire with a single occupied subband should exist at the corner.

To determine the validity of the sharp-corner assumption
in the presence of a $B$-field, the comparison between the magnetic length $l_B$ and the triangular confinement width $W$ becomes relevant.  
The analysis of Ref.~\cite{TEM} is reproduced here, but for the radius of curvature observed in the present sample which is about half that of Ref.~\cite{TEM}.
The  
triangular-well %
wavefunction full-width at half-maximum $W = 18$ nm is estimated from Hartree calculations of the triangular confinement for the above 
typical %
densities. For small magnetic fields such that $W/2 < l_B$ ($B < 8$ T), the $B=0$ Hartree potential can be safely used in place of the finite $B$ Hartree potential, drastically simplifying the dispersion calculation.  At larger magnetic fields such that $W/2 > l_B$ ($B > 8$ T), dispersion calculations should include the $B$-field in the Hartree iteration. Finally, at extreme fields the radius of curvature of the corner $r
= 2.85$ nm becomes important once $r > l_B$ ($B > 81$ T).  At these high fields, the sharp corner approximation for the external potential should be replaced with a real potential with finite corner curvature. The corner-overgrown profiles in Fig \ref{Fig1} are sharp enough that one is easily able to achieve the fractional filling factor $\nu = 1/3$ for typical densities before reaching this limit.
\begin{figure}
\includegraphics[width=\columnwidth]{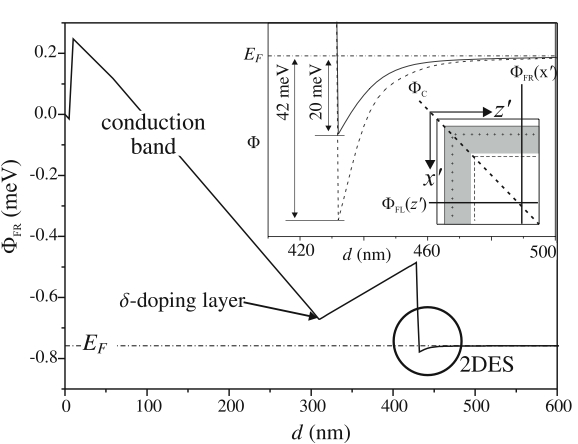}
\caption{Self-consistent $B=0$ Hartree potential ${\Phi}$ for the bent quantum well heterostructure, plotted along a perpendicular cross-section through one of the facets. The two-dimensional electron densities on both facets are \mbox{${1\times 10^{11}\,{\rm cm}^{-2}}$}.
The inset figure shows that the diagonal cross-section through the corner has almost exactly twice the confinement potential as the perpendicular cross-section far from the corner. A 1D accumulation wire results from the additional depth of the potential at the corner.}
\label{zeroB_potential}
\end{figure}
\section{Hartree simulation at $B=0$}
The Hartree simulation of the bent quantum well is presented first at zero magnetic field, assuming a perfectly symmetric sample with equal spacer thicknesses and donor concentrations on both facets yielding a sheet electron density of \mbox{$1.0\times{10}^{11}\,{\rm cm}^{-2}$}. The potential and charge density distribution as well as eigenstates and energy eigenvalues for electrons in the heterostructure are obtained from a self-consistent solution to the Schroedinger equation 
\begin{equation}
\left[\frac{(\vec{p})^2}{2m^{\star}}+\Phi _{cb}(\vec{r})+\Phi _{el} (\vec{r})\right]\Psi(\vec{r})=E\Psi (\vec{r}),
\label{eq:manyElectron}
\end{equation}
and the Poisson equation
\begin{equation}
\Delta \Phi=-\frac{\varrho}{\varepsilon}.
\label{eq:poisson}
\end{equation}
where $\vec{r}=(x,y,z)$ is the position vector, $\varrho$ is the charge density, and $\varepsilon$ is the low-frequency dielectric constant for AlGaAs or GaAs, respectively. The potential $\Phi_{cb}$ is defined by the conduction band of the intrinsic GaAs/AlGaAs semiconductor crystal 
as well as ionized dopants, %
and $\Phi_{el}$ is the Hartree potential of the electron distribution 
solved for overall charge neutrality. %
In the following the total potential $\Phi = \Phi _{cb}+\Phi_{el}$ will be referred to as the Hartree potential.
The Poisson equation is solved for a finite cross-section of the heterostructure with the boundary conditions
\begin{align*}
\Phi_{surf}&=0,\\
\vec{{\cal E}}_{sub}&=0,
\end{align*}
where the 
conduction band %
potential $\Phi_{surf}$ at the sample surface and the electric field $\vec{{\cal E}}_{sub}$ at the interface to the bulk of the sample are set to zero. This surface potential assumes mid-gap pinning of the cap layer.

The Hartree potential $\Phi$ obtained from the self-consistent solution to Eqs. (\ref{eq:manyElectron}) and (\ref{eq:poisson}) is plotted in Fig.\,\ref{zeroB_potential} along a perpendicular cross-section through one facet far from the corner as a function of the distance $d$ from the sample surface. The conduction band bending at the \mbox{GaAs/AlGaAs} interface defines an approximately triangular quantum well which hosts the 2DES in the facets. The inset shows the potential $\Phi_{S}\,(z')$ ($\Phi_{S}\,(x')$), along a perpendicular cross-section through the left (right) facet far from the corner (solid line), and the potential $\Phi_c$ along a diagonal cross-section through the corner junction (dotted line). Due to the confinement in both $x'$- and $z'$-direction the corner potential $\Phi_c$ has about twice the binding potential compared to the facets. The additional depth of the potential well at the corner causes the 1D accumulation with a binding energy of \mbox{7 meV}, compared to \mbox{3.5 meV} for electrons in the 2D facets.
\begin{figure}
\includegraphics[width=\columnwidth]{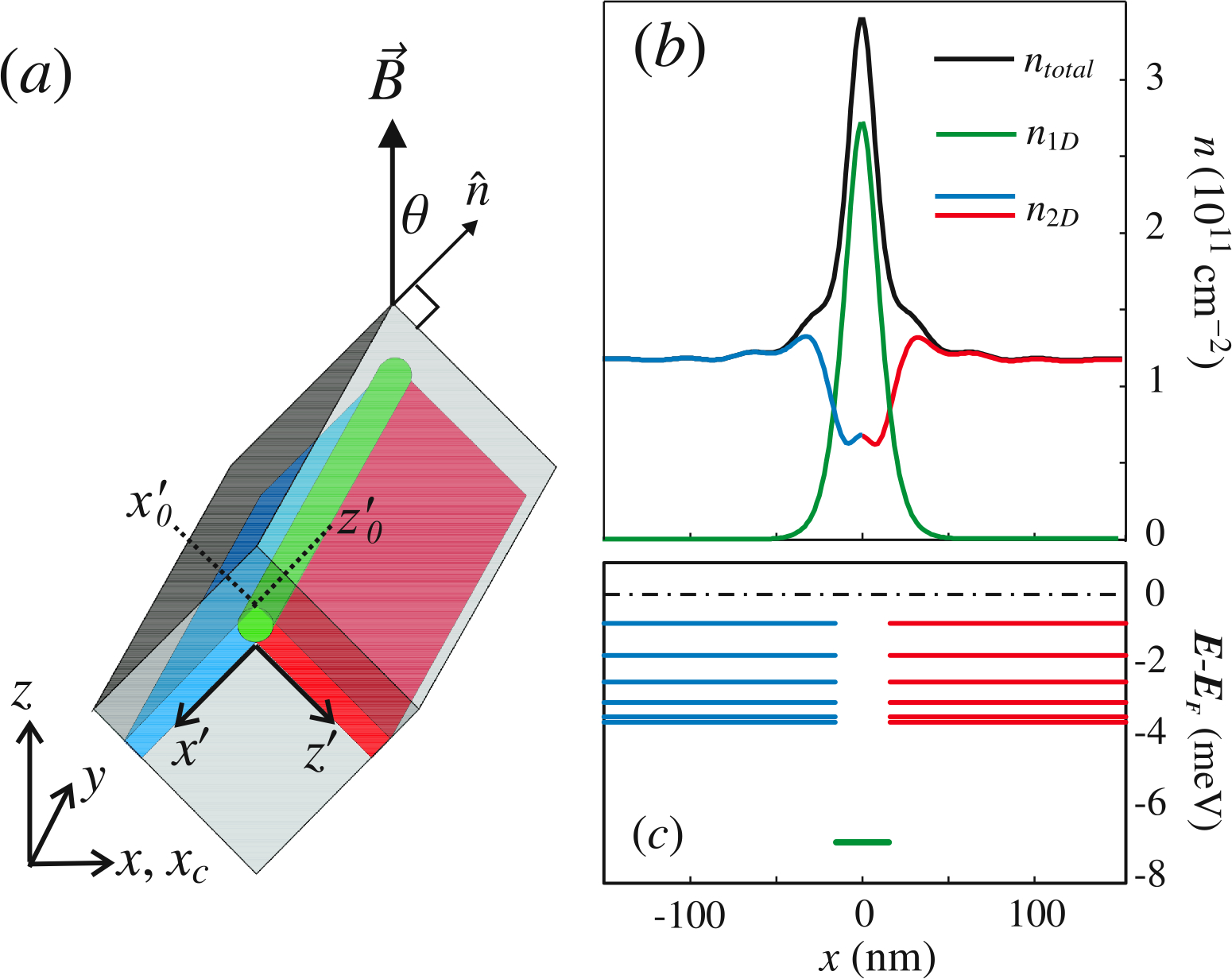}
\caption{(Color online) Panel (a) shows the two coordinate frames defined for the Hartree simulations at finite $B$. The $(x_c,y,z)$-frame is used for the $k_y$-momentum and $x_c$-cyclotron orbit center coordinates, with the z-axis parallel to $\vec{B}$. The $(x',z')$ coordinate frame fixed to the sample is convenient to plot the electronic wavefunctions in real-space. The plots in panel (b) show the calculated electron densities projected onto the x-axis indicated in the schematic, where the blue and red curves are the 2D electron densities in the facets and the blue and red lines at the bottom represent eigenenergies of states in the facets. The green curve is the density of the 1D accumulation wire with eigenenergy represented by the green line at the bottom of the figure. The black curve shows the total electron density. Panel (c) displays the calculated binding energies for the 2D and 1D states.}
\label{Fig2}
\end{figure}
\begin{figure}
\includegraphics[width=0.9\columnwidth]{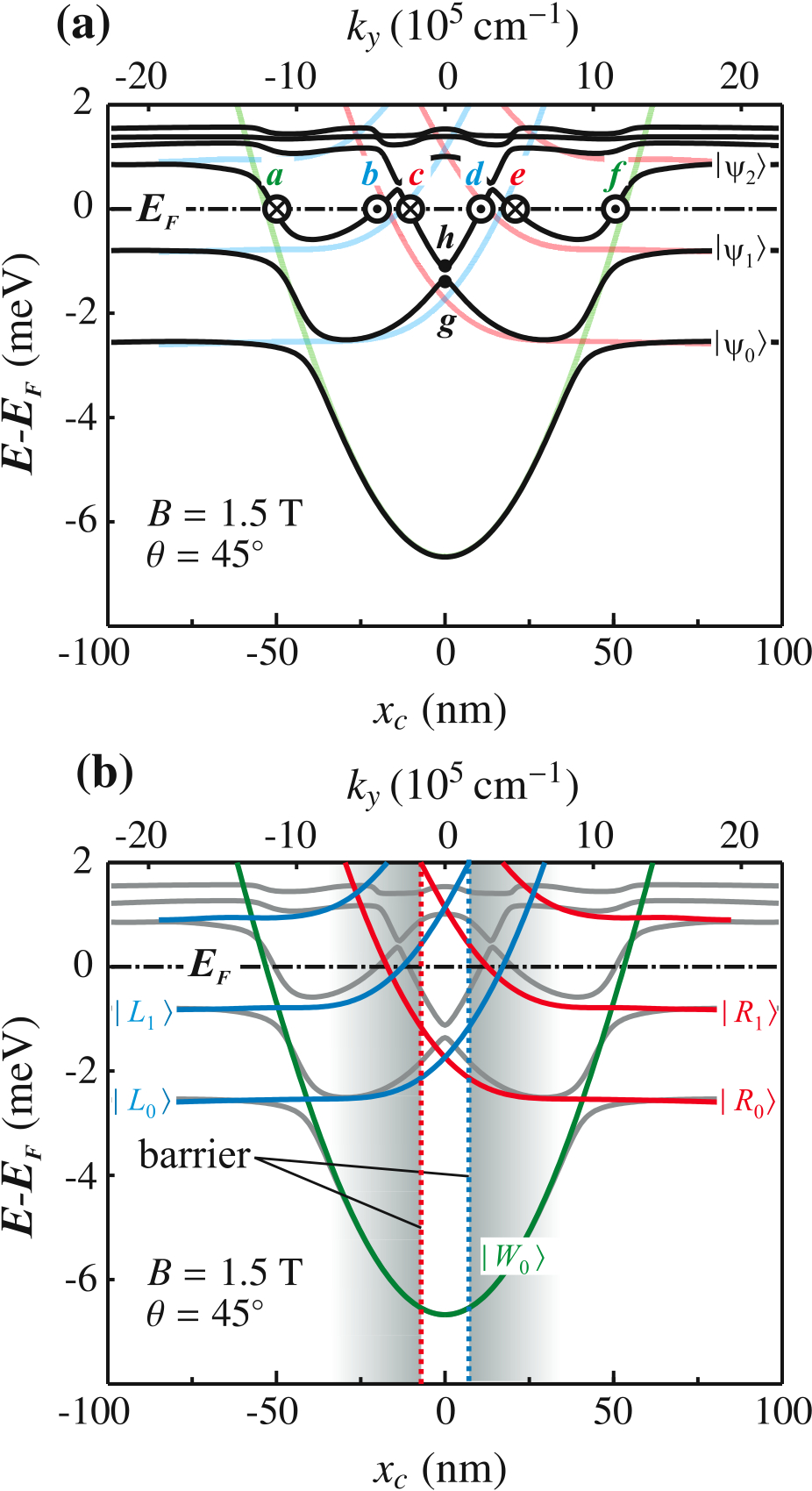}
\caption{(Color online) (a) Hartree dispersions $E$ vs. cyclotron orbit center $x_c$ (momentum $k_y$) corresponding to the wavefunctions $|\psi_0\bigr>$, $|\psi_1\bigr>$, $|\psi_2\bigr>$ (black) of the bent quantum Hall system.   Magnetic field is \mbox{$1.5$ \rm{T}} at a tilt angle $\theta=45^{\circ}$. The six Fermi points responsible for conduction are labelled $a$ through $f$ and are discussed further in Figs.~\ref{Fig6} and \ref{Fig4}.  The anticrossing at $x_c = 0$ has its bonding and anti-bonding states labelled $g$ and $h$, respectively, and is discussed further in Fig.~\ref{Fig5}. (b) The subsystems model of an idealized sharp QH edge.  Dispersions for the left-facet $|L_n\bigr>$ (blue) and right-facet $|R_n\bigr>$ (red) subsystems, and the parabolic dispersion of the 1D wire ground state $|W_0\bigr>$ (green) subsystem are shown. The dotted lines and shaded areas indicate the effective orbit center position of the corner walls as seen by the edge states. For easy comparison of the Hartree solution to the subsystems model, each solutions is plotted in half-tone in the background of the other plot. Wavefunctions for each dispersion branch are labelled according to the discussion in Section IV.}
\label{Fig3}
\end{figure}
Fig.\,\ref{Fig2} (b) shows the electron density calculated at $B=0$, integrated over the quantum well thickness and projected onto the $x_c$-axis indicated in panel (a). We distinguish three different electron systems in the corner region: The 2D systems in the facets plotted in red or blue, respectively, and a 1D accumulation wire plotted in green. Panel (c) shows the calculated binding energies relative to the Fermi level $E_F$ for 2D and 1D states in the bent quantum well, where the 1D wire is approximately twice as deeply bound as the 2D ground energies, due to 
double %
confinement in $z'$-and $x'$-directions. Note that the discrete spectrum of 2D states is an artifact of the finite size of the quantum mechanical simulation.

\section{Hartree simulation at finite $B$}
Having identified the 2D and 1D states that exist in a bent quantum well at zero magnetic field,
we now study the new states that emerge from these subsystems at finite $B$. If a tilted magnetic field is applied such that the 2D electron systems of both facets are in a quantum Hall state, all three subsystems are strongly coupled and form the bent quantum Hall system. In this paper we restrict ourselves to BQH systems where the magnetic field is applied at a $45^{\circ}$-angle such that the 2D systems on both facets are in the same quantum Hall state with filling factor $\nu$.
Zeeman spin splitting and exchange interactions are neglected.  

It has been previously shown that the 
electron screening charge density at %
zero $B$ serves as an excellent approximation to the charge density at finite $B$ \cite{Chklovskii}.  Any small deviations in charge distribution come only from the possible existence of dipolar strips at the boundary between compressible and incompressible strips which are themselves a consequence of screening of Landau-quantized electron density of states in a slowly varying edge potential.  Edge tunneling experiments by Huber, {\it et al.} \cite{Michi} have shown that abrupt quantum Hall boundaries with sharp potentials have vanishingly small incompressible strips.  Thus the zero $B$ Hartree potential will be assumed to serve as an excellent approximation to the finite $B$ Hartree potential.  The presence of a magnetic field will simply add harmonic cyclotron-motion to the total Hamiltonian, derived for completeness below.

It is convenient to pick a coordinate frame $(x,y,z)$, where the $z$-axis is defined by the magnetic field and the $y$-axis is the translationally invariant direction along the corner of the bent quantum well, as shown in Fig.\,\ref{Fig2} (a).
We choose the vector potential $\vec{A}$ in the Landau gauge to preserve $y$-translational invariance and make momentum $k_y$ a good 
quantum number:
\begin{equation}
              \vec{A} =xB\hat{y},
\label{eq:VectPot}
\end{equation}
which satisfies the condition
\begin{equation}
\vec{B} =\vec{\nabla}\times\vec{A}=B\hat{z}
\end{equation}
As an Ansatz for the wavefunctions, we choose
\begin{equation}
\Psi (\vec{r})=\psi _{n,k_y}(x,z)e^{ik_yy}
\end{equation}
or identically
\begin{equation}
\Psi (\vec{r})=\psi _{n,x_c}(x,z)e^{ix_cy/l_B^2}
\label{eq:Ansatz}
\end{equation}
where the cyclotron orbit center $x_c$ 
is proportional to $y$-momentum $x_c=k_yl_B^2$ where \begin{math}l_B=\sqrt{\frac{\hbar}{eB}}\end{math} is the magnetic length, and $n$ is the energy eigenvalue index. 
Only the lowest electrostatically confined sub-band is occupied in these heterostructures, 
so $n$ represents the hybridized Landau subband / wire subband index. %
With the ansatz (\ref{eq:Ansatz}) the explicitly $y$-dependent parts can be separated from the Schroedinger equation
\begin{equation}
\left[\frac{(\vec{p}-e\vec{A})^2}{2m^{\star}}+\Phi(x,z)\right]\Psi(\vec{r})=E\Psi (\vec{r}),
\end{equation}
and one obtains
\begin{eqnarray} 
  \left[\frac{(p_x^2+p_z^2)}{2m^{\star}}+\frac{1}{2}m^{\star}\omega _c^2(x-x_c)^2 +\Phi (x,z)\right] \psi _{n,x_c}(x,z)  \notag
  \\ =E_n(x_c)\psi _{n,x_c}(x,z).
\label{eq:SchrodiB}
\end{eqnarray}
We see that the magnetic field causes the additional harmonic potential $\frac{1}{2}m^{\star}\omega_c^2(x-x_c)^2$ in the $x$-direction perpendicular to $\vec{B}$,
where $\omega _c = \frac{eB}{m^{\star}}$ is the cyclotron frequency. 

%
Fig.\,\ref{Fig3} (a) shows the dispersion calculated for a bent quantum well with sheet electron densities $n=1.0\times10^{11}\,{\rm cm}^{-2}$ in both facets, where a magnetic field of \mbox{$1.5$ T} applied at a tilt angle $\theta=45^{\circ}$ relative to the facet normal vectors brings both 2D systems to the $\nu=4$ quantum Hall state. The dispersion of this bent quantum Hall system is plotted in black as energy $E$ versus the momentum $k_y$ (top axis) and the orbit center $x_c$ (bottom axis).

\section{Subsystems model}

To provide intuition as to the origin of these states, we define three separate subsystems which hybridize to form the bent quantum well system, namely the left facet quantum well, the right facet quantum well, and the 1D accumulation wire at the corner.  All subsystems are 1D in character, as their energy is uniquely determined in terms of a single spatial coordinate, the cyclotron orbit center $x_c$. We will show that not only the overall dispersion, but also the wavefunctions themselves can be quantitatively predicted from such a subsystems model.  The candidate wavefunctions at a given orbit center coordinate are generated by taking the subsystem wavefunction at the desired orbit center and energy and projecting out all more deeply bound states at the same orbit center to form an orthogonal basis.

\subsection{Subsystem dispersions}

The separate left-facet and right-facet quantum well subsystems are modeled as heterojunction triangular-well potentials which terminate abruptly in a \mbox{$0.3$ eV} hard wall at the corner, defined in Fig.~\ref{Fig2}(a) as a wall at $x'_0$ for the left facet and at $z'_0$ for the right facet. Note that the hard wall boundary perceived by a given quantum well arises because of the sudden $90^{\circ}$ bend in the heterojunction.  The result is a pair of coupled orthogonally facing hard-wall-like QH edge dispersions. When projected onto a common axis, such as $x$, the finite width of the wavefunction results in different projected positions of the effective hard wall for the opposing dispersions.    

The triangular well potential is taken to be the Hartree solution of the quantum well far from the junction $\Phi_S$ as calculated in Section III.
%
\begin{align}
\Phi_{L}(x',z')= \begin{cases}
\Phi_B&,\,\,\,\, x'<x'_0\\
\Phi_S(z')&,\,\,\,\, x'>x'_0\\
\end{cases}\\
\Phi_{R}(x',z')= \begin{cases}
\Phi_B&,\,\,\,\, z'<z'_0\\
\Phi_S(x')&,\,\,\,\, z'>z'_0\\
\end{cases}
\label{pot_definition}
\end{align}
and \mbox{$\Phi_B=E_F+0.3$ eV} 
is used to approximate the corner heterojunction barrier. %

The confinement potential $\Phi_W(x',z')$ of the 1D accumulation wire is modeled as the 
sum of two orthogonal triangular potentials %
\begin{equation}
\Phi_W(x',z')=\Phi_{S}(x') + \Phi_{S}(z').
\end{equation}

In analogy to Eq. (\ref{eq:SchrodiB}) for the Hartree simulation, the left, right, and wire subsystems satisfy the following Schroedinger equations:
\begin{eqnarray} 
  \left[\frac{(p_x^2+p_z^2)}{2m^{\star}}+\frac{1}{2}m^{\star}\omega _c^2(x-x_c)^2 +\Phi_L\right] L _{n,x_c}(x,z) \notag
  \\ =E^L_n(x_c)L _{n,x_c}(x,z)
\label{eq:psiL}
\end{eqnarray}
\begin{eqnarray} 
  \left[\frac{(p_x^2+p_z^2)}{2m^{\star}}+\frac{1}{2}m^{\star}\omega _c^2(x-x_c)^2 +\Phi_R\right] R _{n,x_c}(x,z) \notag
  \\ =E^R_n(x_c)R _{n,x_c}(x,z)
\label{eq:psiR}
\end{eqnarray}\begin{eqnarray} 
  \left[\frac{(p_x^2+p_z^2)}{2m^{\star}}+\frac{1}{2}m^{\star}\omega _c^2(x-x_c)^2 +\Phi_W\right] W _{n,x_c}(x,z) \notag
  \\ =E^W_n(x_c)W _{n,x_c}(x,z)
\label{eq:psiW}
\end{eqnarray}

The dispersion calculated for the left-facet quantum well is shown in blue in \mbox{Fig.\,\ref{Fig3} (b)}, mirrored by the red dispersion of the right-facet quantum well at \mbox{$B=1.5$ T}, $\nu=4$.
 Effective positions for the hard wall seen by either facet is indicated by vertical dotted lines, offset from the geometric center by the projected finite width of the wavefunction.
The deeply bound accumulation wire is indicated by the green parabolic dispersion in Fig.\,\ref{Fig2}. 

The dispersions of these constituent systems in Fig.~\ref{Fig3}(b) clearly represent the key features of the total Hartree dispersion shown in Fig.~\ref{Fig3}(a).
First, the low-energy dispersion of the 1D subsystem accumulation wire perfectly matches the dispersion of the full Hartree calculation at low energies.  The quantum Hall edge subsystems also parallel the branches of the full Hartree calculation, though the energies of the subsystem edge states are slightly lower.   This may be a result of level repulsion which is not taken into account in the subsystems model.  %

Coupling between the different subsystems arises because of interpenetration of the wavefunctions at the corner.  Semiclassically, one can consider an electron at low magnetic fields spending most of its time executing the arc of its cyclotron orbit within the facet, but for a short time perturbed by states at the corner as it reflects specularly from the hard wall.  Quantum mechanically, the interpenetration of the counter-propagating edge states from the two orthogonal facets and the accumulation wire at the corner, gives rise to anticrossing gaps at all intersections of the subsystem dispersions. States with larger spatial overlap integral (such as the wire and facet states) have larger coupling and larger anticrossing gaps, whereas states with less spatial overlap (such as between opposing facets) have smaller anticrossing gaps.  The total dispersion of the bent quantum Hall system emerges as a hybrid system of these three coupled 1D subsystem dispersions along the corner. 

The states which are responsible for determining electron conduction along the corner are those where the dispersion curves cross the Fermi level $E_F$, highlighted in Fig.\,\ref{Fig3} (a) and labelled $a$ through $f$.
The direction of propagation of each mode is determined by the slope of the dispersion at $E_F$, where a positive slope leads to a forward-propagating mode and a negative slope yields a reverse-propagating mode. The left-facet edge states $b$ and $d$ (blue) as well as the wire state $f$ are forward-propagating, and the right-facet edge states $c$ and $e$ (red) and wire state $a$ are reverse-propagating. 
At low magnetic fields and integer filling factor such that the facets are gapped, such as $\nu = 4$ in Fig.\,\ref{Fig3} (a), the Hartree simulations show $\nu$ spin-degenerate conducting modes from each quantum Hall edge. The 1D accumulation wire adds two spin-degenerate modes in each direction and serves as an additional 1D channel for scattering. Together with the 1D edge modes from the QH systems, the model thus predicts up to $N = \nu + 2$ one-dimensional modes in each direction. 

\begin{figure}
\centerline{\includegraphics[width=\columnwidth]{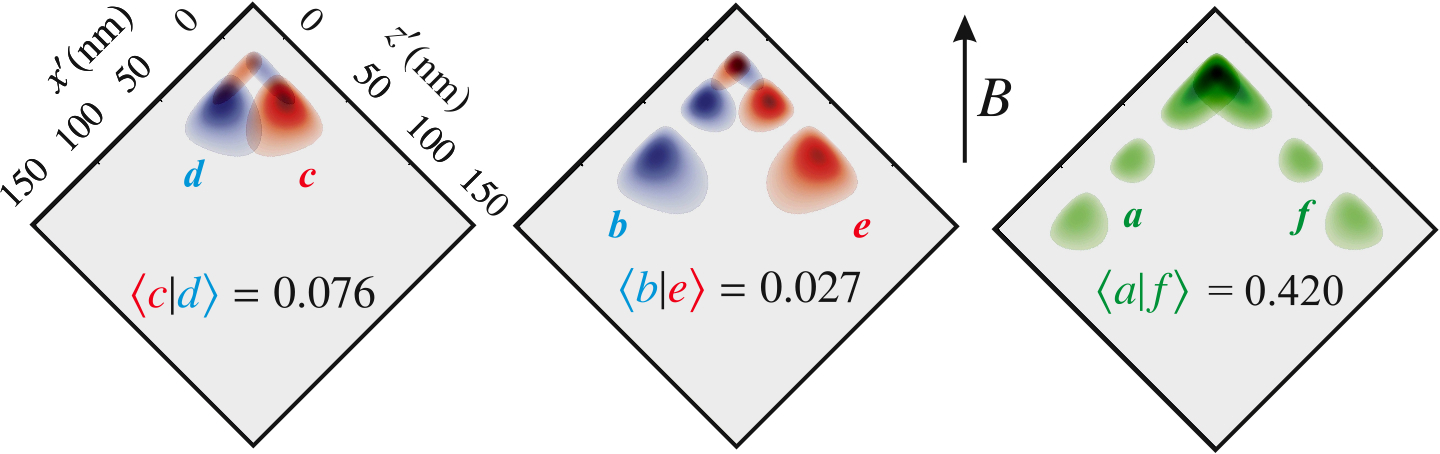}}
\caption{(Color online) Color-map plots of the probability densities for real-space wavefunctions corresponding to the various 1D modes marked in Fig.\,\ref{Fig3} (a). The electrons in the facets (blue and red) and corner accumulation wire (green) are colored according to their Hartree dispersions shown in Fig.\,\ref{Fig3} (b).}
\label{Fig6}
\end{figure}

Fig.\,\ref{Fig6} shows the real-space wavefunctions of these same 1D modes $a$ through $f$ from Fig.\,\ref{Fig3} (a). The sharp edge potential allows for a finite overlap of the wavefunctions, so that any disorder potential can scatter charge between dispersion branches, including backscattering.  Overlap integrals between different backscattering states at $E_F$ are calculated, ranging from the smallest overlap $\bigl<b|e\bigr> = 0.027$, to a typical value $\bigl<c|d\bigr> = 0.076$, to the largest overlap integral for backscattering states $\bigl<a|f\bigr> = 0.420$ which is found inside the wire.   One therefore does not expect to observe universal conductance values for the corner conductance, but instead to measure non-universal values which characterize scattering along such a corner \cite{4}.  

%

\begin{figure}
\includegraphics[width=\columnwidth]{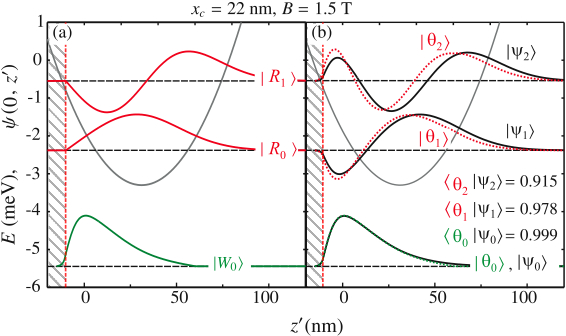}
\caption{(Color online) (a) Crossections along the $z'$-axis at $x'=0$ of the right quantum Hall edge subsystem wavefunctions $|R_0\bigr>$, $|R_1\bigr>$ and the 1D accumulation wire subsystem $|W_0\bigr>$ at \mbox{$B=1.5$ T} and $x_c=22\,{\rm nm}$.  (b) Candidate wavefunctions $|\theta_1\bigr>$ and $|\theta_2\bigr>$ for the combined bent quantum well system are constructed by projecting out the deeply bound wire state $|W_0\bigr>$ from the QH edge subsystem according to Eqs. \ref{eq:project1} and \ref{eq:project2}. Together with the wire ground state these states show excellent agreement with the Hartree eigenstates $|\psi_0\bigr>$, $|\psi_1\bigr>$, and $|\psi_2\bigr>$ at $x_c=22\,{\rm nm}$ obtained from the numerical simulation of the complete bent quantum Hall system (inset equations), confirming the correspondence with the subsystems model.}
\label{Fig4}
\end{figure}

\subsection{Subsystem wavefunctions}
In this section we show 
how the wavefunctions of the total hybrid system can be understood in terms of the wavefunctions of the individual subsystem eigenstates.  We begin by examining states away from the anticrossings, and will define trial states $|\theta_n\bigr>$  in terms of the subsystem states $|L_n\bigr>$, $|R_n\bigr>$, and $|W_n\bigr>$, all with the same orbit center $x_c$.  The explicit functional dependence on the orbit center $x_c$ will be dropped
hereafter, since all wavefunctions must have the same $x_c$
value to constitute $k_y$-momentum eigenstates.
%


To start, consider an orbit center $0 < x_c < 40\,{\rm nm}$ where the wire has the lowest ground state energy.  Within this range, the state $|W_0\bigr>$ at the bottom of Fig.~\ref{Fig4}(a) is a good trial for the proper Hartree ground state $|\psi_0\bigr>$, leading us to the ground state ansatz $|\theta_0\bigr>$. 

\begin{equation}
|\theta_0\bigr>=|W_0\bigr>
\label{eq:project0}
\end{equation}

\noindent The case for $x_c=22\,{\rm nm}$ is plotted in the bottom of Fig.~\ref{Fig4}(b).  The overlap between this trial state and the Hartree solution shows excellent agreement:

\begin{equation}
\bigl<\psi_0|\theta_0\bigr> = 0.999
\end{equation}

The next higher subsystem state at this $x_c$ is $|R_0\bigr>$ shown in the middle of Fig.~\ref{Fig4}(a).  This state is not orthogonal to the ground state $|\theta_0\bigr>$, so the trial state for our first excited energy $|\theta_1\bigr>$ requires that we project out this ground state:

\begin{equation}
|\theta_1\bigr> = {N}_1 \left\{|R_0\bigr> - |\theta_0\bigr>\bigl<\theta_0|R_0\bigr>\right\}
\label{eq:project1}
\end{equation}

\noindent where $N_1$ is a normalization constant.  The resulting wavefunction is shown in the middle of Fig.~\ref{Fig4}(b) and overlaps very well with the Hartree solution $|\psi_1\bigr> $

\begin{equation}
\bigl<\psi_1|\theta_1\bigr> = 0.978
\end{equation}

Continuing to the next higher energy wavefunction, the state $|R_1\bigr>$ shown at the top of Fig.~\ref{Fig4}(a) needs to be made orthogonal to all states $|\theta_1\bigr>$ and $|\theta_0\bigr>$ below it, thereby generating the trial state $|\theta_2\bigr>$:

\begin{align}
|\theta_2\bigr> = {N}_2 \bigl\{|R_1\bigr> - |\theta_0\bigr>\bigl<\theta_0|R_1\bigr>\notag  \\
-|\theta_1\bigr>\bigl<\theta_1|R_1\bigr>\bigr\}
\label{eq:project2}
\end{align}

\noindent Again the result strongly overlaps with the Hartree solution $|\psi_2\bigr> $

\begin{equation}
\bigl<\psi_2|\theta_2\bigr> = 0.915
\end{equation}

In general, if one defines the energetically ordered sequence of subsystem states at a given $x_c$ called $|S_n\bigr>$, which for this example was $|S_0\bigr> = |W_0\bigr>$, $|S_1\bigr> = |R_0\bigr>$, $|S_2\bigr> = |R_1\bigr> ...$, then trial wavefunctions can be generated from the expression below:

\begin{equation}
|\theta_n\bigr> =N_n\left\{\mathbbm{1}-\sum\limits_{i=0...n-1}|\theta_i\bigr>\bigl<\theta_i|\right\}|S_{n}\bigr>
\label{eq:project3}
\end{equation}

\noindent These trial states confirm that the low energy Hartree eigenstates for the total corner quantum well can be understood in terms of the subsystem wavefunctions.

\begin{figure*}
\centerline{\includegraphics[width=\textwidth]{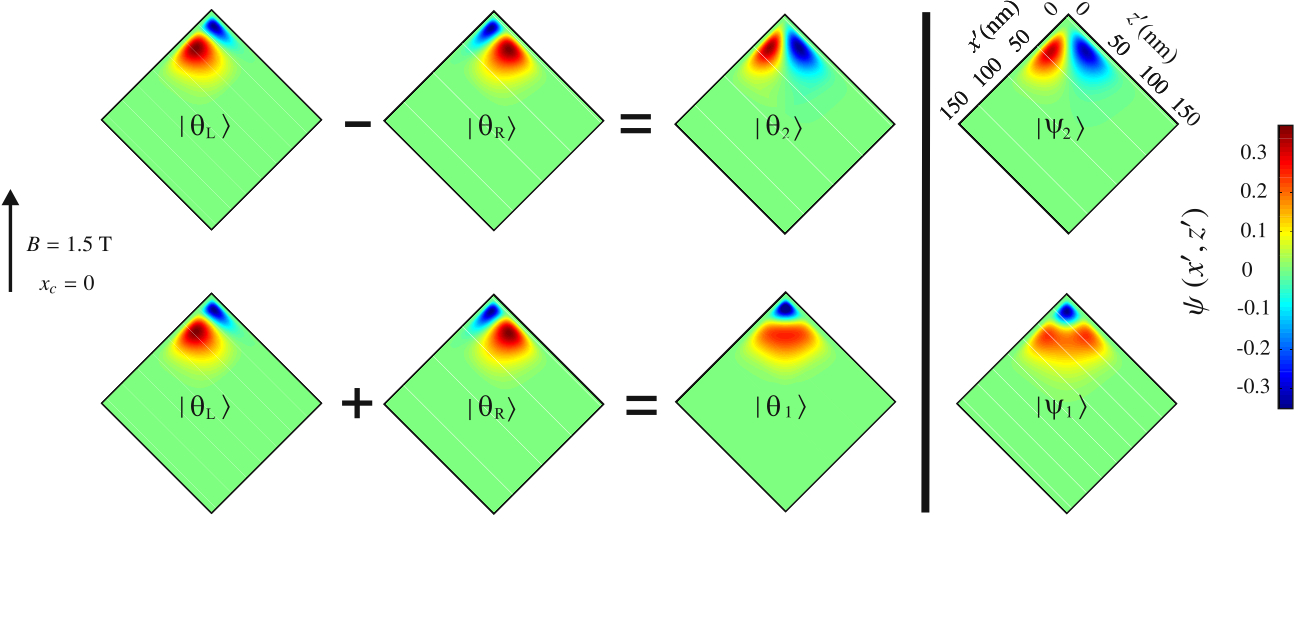}}
\caption{(Color online) Candidate wavefunctions for anticrossing points $g$ and $h$ in the dispersion of Fig.~\ref{Fig3} are shown as color scale probability amplitudes.
Energy degenerate trial states $|\theta_L\bigr>$ and $|\theta_R\bigr>$ are plotted, along with their symmetric and antisymmetric combinations $|\theta_1\bigr>$ and $|\theta_2\bigr>$, respectively.  
The resulting wavefunctions agree very well with the numerical Hartree results $|\psi_1\bigr>$ for point $g$ and $|\psi_2\bigr>$ for point $h$, with an overlap integral of 97\% - 98\%.}
\label{Fig5}
\end{figure*}

Finally, we examine the wavefunctions at an anticrossing such as $x_c = 0$ to check how the above expression for trial states changes in the case of degeneracy.  For anticrossing states in Fig.~\ref{Fig3}(a) designated as points $g$ and $h$, the subsystem states $|L_0\bigr>$ and $|R_0\bigr>$ are degenerate, so the sequence of states $|S_n\bigr>$ needs to be modified to consider first the symmetric, and then the antisymmetric combination:  $|S_0\bigr> = |W_0\bigr>$, $|S_1\bigr> = |R_0\bigr>+|L_0\bigr>$, $|S_2\bigr> = |R_1\bigr>-|L_0\bigr>$.  Fig.\,\ref{Fig5} shows color map plots of the wavefunctions $\psi_n(x',z')$ corresponding to states at the $x_c=0$ anti-crossing of the left and right quantum Hall edges. The states  $|\theta_L\bigr>$ and $|\theta_R\bigr>$ are constructed from the wavefunctions $|L_0\bigr>$ and $|R_0\bigr>$ separately, whereas $|\theta_1\bigr>$ and $|\theta_2\bigr>$ take the degeneracy into account, and show excellent agreement (97-98\% overlap) with the Hartree results $|\psi_1\bigr>$ and $|\psi_2\bigr>$ for the binding or anti-binding states at the $x_c=0$ anti-crossing gap.

\section{Hartree simulation at large $B$}
\begin{figure}
\centerline{\includegraphics[width=\columnwidth]{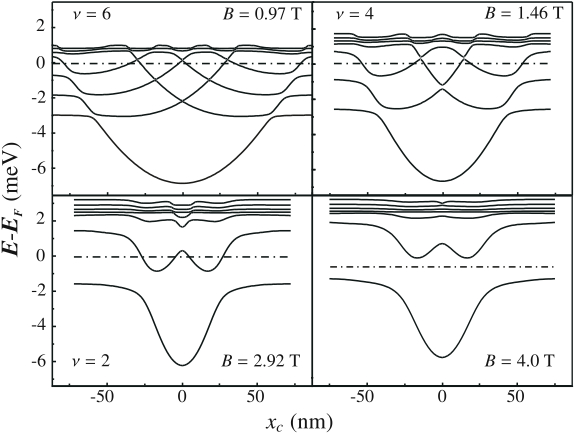}}
\caption[Hartree dispersions of a bent quantum Hall junction at various magnetic fields.]{Hartree dispersions of a bent quantum Hall junction at various magnetic fields. In contrast to planar barrier systems where the gaps vanish exponentially at high $B$, here the anti-crossing gaps due to strong coupling of the modes at the corner {\em increase} with increasing magnetic field.}
\label{Fig7}
\end{figure}
Fig.\,\ref{Fig7} shows Hartree dispersions of the bent QH system at various magnetic fields. With increasing $B$ the anti-crossing gaps in the dispersion increase, indicating an increasing coupling among the three constituent subsystems. This follows from the reduced magnetic length, which brings all quantum confined 1D states closer to the corner, and therefore in stronger overlap with each other. This is in contrast to planar barrier systems, for example the planar antiwire of Refs.\,\cite{81,82}, where the coupling {\it decreases} exponentially at high $B$.
The strong anticrossings result in both positive curvatures, corresponding to an electron-like mass, and negative curvatures, representing a hole-like mass, in various parts of the dispersion.  The band gaps between hybridized Landau bands start to become larger than the band-widths of the higher excited bands. The signature feature of a bound wire state at the corner evolves at high magnetic fields into a central dispersion minimum for the lowest band.

\section{Interpretation of experiments}

A series of transport experiments have already been performed on bent quantum Hall systems.
Conductance measurements along the BQH junction have shown weakly insulating, strongly insulating and metallic behavior, depending on $\nu$ \cite{4}. 
At low magnetic fields $(\nu = 3,4,5,6)$ weakly insulating behavior was observed, with a finite, yet non-universal conductance of a few percent of the quantized value $\nu e^2/h$ and little temperature- or dc voltage bias dependence. In contrast, the conductance at higher magnetic fields $(\nu = 1,2)$ showed strongly insulating behavior and decreased drastically with decreasing temperature or voltage. 
Finally, at the fractional filling factor $\nu = 1/3$ the BQH junction showed metallic behavior, where the conductance strongly increased upon lowering the temperature or voltage.
Possible explanations for the unique transport properties of the bent quantum Hall junction, can be sought within the microscopic model presented here.

We discuss first the experimental case of integer filling factor $\nu \geq 3$, which was observed to be a  weak insulator \cite{4,thesis}.  The microscopic model for such a system was shown here to be a 1D multimode conductor such as in Fig.\,\ref{Fig3}(a) for the $\nu=4$ dispersion case, with Fermi points for each mode indexed $i = a ... f$.  Note that the experiments of Huber {\em et al.} \cite{Michi} have demonstrated that at a sharp edge potential, the wavefunctions of different co-propagating edge states are close enough to each other to spatially overlap.  In Fig.\,\ref{Fig4} we see a similar scenario, but in addition, the counter-propagating states also overlap.  A disorder potential $V(x',y,z')$ thus allows backscattering of charge among these Fermi points in Fig.\,\ref{Fig3}(a) .   

To understand how to notate momentum scattering in the bent QH system, the Fourier transform of the disorder potential can be taken in the $y$ direction along the length of the corner:
\begin{equation}
V_{k_y} (x',z') = \int dy\,e^{-i  k_yy} \, V (x',y,z')
\end{equation}

\noindent Using the notation of Eq.\,\ref{eq:Ansatz} for the wavefunction, the disorder scattering matrix element between Fermi points $i$ and $j$ can be expressed as:
\begin{equation}
V_{ij} = \bigl< \psi_{n_i,x_i}(x',z')| \, V_{k_{ij}} (x',z') \, |\psi_{n_j,x_j}(x',z')\bigr>
\end{equation}

\noindent where $n_i$ and $x_i$ are the energy subband index and cyclotron orbit center of the $i^{\rm th}$ Fermi point, and $k_{ij} = (x_j-x_i) /l_B^2$ is the momentum scattering wavevector in terms of the orbit center coordinates. The expression can thus be used to describe the experimentally observed non-universal conductance for $\nu \geq 3$ in terms of disorder-induced scattering among various 1D modes.

Next we discuss the experimental case of integer filling factor $\nu = 1, 2$, which was observed to be  strongly insulating.  A detailed analysis of the experimental temperature and voltage dependence has already been performed, and reveals 1D hopping conduction along the corner, with 1D activated hopping at higher temperatures $(T > 320\,{\rm mK})$ giving way to 1D variable range hopping at low temperatures $(T < 320\,{\rm mK})$ \cite{4,thesis}.  We can understand this in terms of the Hartree simulations calculated here in the high-$B$ limit, such as in Fig.\,\ref{Fig7} which shows a band insulator at large $B$. Although band conduction is inhibited since none of the dispersions intersect the Fermi energy, the conventional understanding of the QHE acknowledges that disorder will induce localized states within the gap.  When the Fermi energy sits within these localized states, standard theory of semiconductors \cite{Shklovskii} would predict activated hopping conduction among localized states at higher temperatures, which would give way to variable range hopping within the impurity band at lower temperatures, as seen experimentally.  The intriguing observation is that hopping conduction appears to take place only along the 1D corner junction whereas it is suppressed in the facets.  This could be caused by an enhanced the density of disorder localized states at the corner, which results from dips in the hybridized dispersions at the corner (see bottom-right panel of Fig.\,\ref{Fig7}).  Thus the Hartree calculations are able to provide a plausible microscopic model for the strongly insulating behavior at $\nu = 1, 2$.  


Finally we discuss the experimental case of extreme magnetic fields of approximately \mbox{23 T} needed to reach $\nu=1/3$ in both facets \cite{4}.  A proper electrostatic modeling of this case would require a self-consistent treatment including electron-electron interactions to correctly arrive at the Laughlin ground state in the facets, and conduction of charge along the corner would have to be described in terms of 1/3-charge quasiparticles.  Thus any model proposed here will be speculative in nature.  Nonetheless, even though such numerical calculations are beyond the scope of this work, one can still arrive at a qualitative picture by considering the electrostatics revealed by the single particle picture in the high field limit.  Fig.\,\ref{Fig7} identifies the energy minimum due to the accumulation wire at the center of the corner to be several mV below the ground energy in the facets.  Whereas the ground state far from the corner is expected to be a standard Laughlin state with probability of occupancy 1/3 per state, the deep energetic minimum at the corner will exceed the interaction gap energy and thus be fully occupied with one electron per state.  The corner thus represents a $\nu = 1/3:1:1/3$ junction.  
 
As discussed in Refs.\,\cite{4,Kane,Renn}, coupled fractional QH edges can result in the 1D metallic behavior in the so-called anti-wire geometry. In this system, two counter-propagating $\nu = 1/3$ edges backscatter electron charge through a depeted $\nu = 0$ vacuum.   it is important that electrons (not fractional quasiparticles) backscatter the charge between the counter-propagating $\nu = 1/3$ edges, so that backscattering of charge becomes less and less relevant at decreasing temperatures, thus wire conductivity increases as the temperature drops. Prerequisites for the anti-wire model therefore include a sufficiently strong spatial overlap of counter-propagating edge modes to allow charge scattering, and an effective 'vacuum' for the fractional quasiparticles between edge modes to ensure that electrons, not fractional quasiparticles, are backscattered. The Hartree simulations presented in this paper 
infer that both conditions are fulfilled in the bent quantum Hall system at hand.  The dispersions shown in Fig.\,\ref{Fig7} indicate an increasing coupling of counter-propagating edge states leading to spatial overlap of the wavefunctions, and the 1D accumulation wire in Fig.\,\ref{Fig3} can function as an effective vacuum for the $\nu=1/3$ quasiparticles, since at $\nu = 1$ no fractional excitations can exist in the fully occupied deeply bound wire-region of the dispersion.  Thus the Hartree simulations allow us to construct a candidate picture of the high $B$ field metallic behavior at the corner junction in terms of a quantum Hall anti-wire.

This situation is in contrast to the scenario of the planar anti-wire geometry originally proposed in Refs.\,\cite{4,Renn,Kane} and implemented in Refs.\,\cite{81,82}.  Here the tunnel barrier separating the two coplanar QH systems exponentially suppresses tunneling at high $B$, thereby prohibiting the desired strong coupling of fractional QH edges. We thus propose that, at the high magnetic fields necessary to see $\nu = 1/3$ edges, the bent quantum Hall system may actually be the only experimentally realizable system which can show the predicted 'anti-wire' system of Refs.\,\cite{Renn,Kane}, because the corner geometry allows the strong coupling of two distinct $\nu=1/3$ edges. 

\section{Conclusion}
We have studied a new type of low-dimensional system, the bent quantum Hall junction, in spin-degenerate Hartree simulations, providing an elementary understanding of the various coupled electronic states existing in such a non-planar junction of two quantum Hall systems. We have shown how the electronic dispersion and eigenstates in this hybrid system can still be quantitatively understood in terms of the dispersion and eigenstates of constituent QW facet and accumulation wire subsystems. Such analysis can simplify future simulations: instead of a fully numerical solution to the Schroedinger equation one could solve the 1D Hartree potential of a single quantum well, and then construct the dispersion and eigenstates from template wavefunctions of the subsystem states. With such a semi-analytical approach the simulations could also be extended to allow for self-consistent simulations at finite magnetic fields or to approximate the hybridized fractional QH states expected in the BQH junction at high $B$. The simulations clearly demonstrate how the bent quantum Hall system differs from any tunnel-coupled planar system of counter-propagating QH edges. Particularly the increasing wavefunction overlap of the edge states with increasing $B$ makes it an interesting device to study coupled fractional quantum Hall edges, and possibly the only system where the predicted 1D metallic behavior at $\nu=1/3$ can be observed experimentally.

\begin{acknowledgements}
\noindent This work was supported by the Deutsche Forschungs\-gemeinschaft in the Schwerpunktprogramm Quanten Hall Systeme, by DFG GR 2618/1-1, by NSF CAREER Award DMR-0748856, the Sonderforschungsbereich SFB 621 and the Nanosystems Initiative Munich.
\end{acknowledgements}

\noindent $^{\dagger}$corresponding author: m-grayson@northwestern.edu\\

\clearpage



\end{document}